\documentclass[journal]{IEEEtran}
\ifCLASSINFOpdf
\else
\fi

\usepackage{array}
\usepackage{multirow}
\usepackage{amsmath}
\usepackage{amssymb}
\usepackage{graphicx}
\usepackage{xurl}

\begin{document}
\title{Obtaining perceptually equivalent resolutions in handheld devices for streaming bandwidth saving}

\author{Mateo C\'amara, C\'esar D\'iaz, Juan Casal, Jorge Ruano and Narciso Garc\'ia}% <-this % stops a space
\thanks{M. C\'amara, C. D\'iaz and N. Garc\'ia are with the Grupo de Tratamiento de Im\'agenes, Information Processing and Telecommunications Center and ETSI Telecomunicaci\'on, Universidad Polit\'ecnica de Madrid, 28040 Madrid, Spain (e-mail: mcl@gti.ssr.upm.es; cdm@gti.ssr.upm.es; narciso@gti.ssr.upm.es)}% 
\thanks{J. Casal and J. Ruano are with Nokia, Mar\'ia Tubau 9, 28050 Madrid, Spain (e-mail: juan.casal\_martin@nokia.com; jorge.ruano\_puente@nokia.com)}%
\thanks{Manuscript received xxxx, 2018; revised xxxx, 2018.}

\markboth{Journal of \LaTeX\ Class Files,~Vol.~14, No.~8, August~2015}%
{Shell \MakeLowercase{\textit{et al.}}: Bare Demo of IEEEtran.cls for IEEE Journals}

\maketitle

\begin{abstract}
We present the description and analysis test conducted to find whether  is looking for savings in bandwidth consumption of audiovisual content without reducing its quality. This report shows the results, as well as the work guidelines and the dynamics that have been followed. The objective that is proposed for this project is to seek the quality limit for which a user stops perceiving an increase of some parameter as an improvement. In particular, we are looking forward to know the limit resolution, trying to isolate as much as possible other parameters such as bitrate or packet loss. In this sense, subjective tests have been carried out on 50 subjects, who have provided a total of 5 000 scores. Assessments also were useful to know about human biases, for example memory or fatigue. Therefore, the conclusions that can be drawn are not only favorable for mathematical and / or subjective criteria, but also for the own experience that the researchers have had when dealing with subjects of very different characteristics that provided additional personal feedback to the tests carried out.
\end{abstract}


\IEEEpeerreviewmaketitle

\section{Introduction}
\label{sec:introduction}

\IEEEPARstart{V}{ideo} traffic accounts for more than 75\% of all IP traffic today, and this percentage is expected to grow noticeably over the next years. Furthermore, smartphone traffic will reach approximately one third of total IP traffic by 2021~\cite{2017-Cisco}. Additionally, despite the appearance in the last few years of new significantly more efficient video compression standards like H.264/MPEG-4 Part 10 Advanced Video Coding (AVC)~\cite{2003-Wiegand}, H.265/MPEG-H Part 2 High Efficiency Video Coding (HEVC)~\cite{2012-Sullivan} or VP9~\cite{2016-Grange}, the bitrates required to provide sufficiently good quality content to end users are still very high~\cite{2018-Apple}. Faced with these overwhelming data, service and content providers are constantly searching for means to supply multimedia content at the same or better quality with the lowest bitrate possible.

Typically, the initial approach implemented to generate content at different qualities is the use of general purpose fixed encoding ladders~\cite{2018-Apple}. These ladders are tables that explicitely indicate the values of several parameters (e.g. image resolution, framerate, bitrate...) to create a given level of quality: 4K Ultra High Definition (4K UHD), Full HD, HD Ready, etc. Nevertheless, this approach presents important drawbacks that limit the search for ways to save bandwidth, like wasting bitrate in encoding content of low complexity (e.g. news, cartoons, sitcoms...) or not using the necessary for high complexity content (e.g. soccer, action movies...). These disadvantages can however be limited by using more flexible ladders that allow to specify ranges of bitrates instead of predetermined specific values for every layer~\cite{2018-YouTube}. The performance of these strategies can be further improved by means of using not one but several ladders, each one of them focused on a particular type of content. This per-category encoding approach is however insufficient, as sequences classified into the same category might not be of comparable level of complexity. This is why Netflix pioneered in late 2015 the implementation of per-title encoding, i.e., using different sets of encoding settings per sequence~\cite{2015-Netflix}. Per-title encoding schemes evaluate the complexity of each video and create a specific encoding ladder for it. But once again, this might not be enough, as the complexity of videos may change notably over time. Thus, new approaches based on either per-shot or per-chunk encoding schemes are being developed today to better adapt the encoding settings to the actual complexity of the content that is to be encoded, segmented and streamed to the users. [ALGO SOBRE C.E.]

Additionally, most of these content-driven frameworks also enable device-aware encoding to further boost efficiency. This approach enables the division of ladders into per-device subladders. These subladders present encoding settings adjusted to the characteristics of the targeted type of device. The majority basically use the same target quality levels in all subladders and simply adapt the value of the encoding parameters according to the type of device in question or directly remove some of these levels. This is done, for instance, not including substandard or standard definition levels for Smart TVs or 4K-UHD for mobile phones. So far, this has been done rather automatically, that is, without considering the actual specifications of the devices and how they impact the quality actually perceived by users. In this respect, it is desirable to introduce a perceptually-driven preparation of device-aware encoding ladders to properly improve efficiency.

In this respect, this work describes the design, development and analysis of subjective assessment tests to find the so-called equivalent quality actually perceived by users in handheld devices. The equivalent quality is defined as the quality level from which users do not statistically significantly perceive an improvement if objectively better quality sequences are presented to them. In this way, it is possible to know to what extent it is relevant (or even necessary) to improve, or not, the quality of the sequence provided to the user. In particular, this work is focused on finding the maximum resolution that users actually perceive depending on the characteristics of the handheld device and the type of content that is delivered. The bitrate or other quality-related criteria are beyond the scope of this set of tests.

The paper is structured as follows. In Section \ref{sec:related_work}
we describe the most significant works related to this very topic.
In Section \ref{sec:test_features} we present all the features of the conducted subjective assessment. Then, we present and analyze the results of the tests in Section \ref{sec:results}. Finally, the conclusions are included in Section \ref{sec:conclusion}.

\section{Related work}
\label{sec:related_work}

Nevertheless, several studies have been consulted. In the paper of~\cite{2017-Brunnstrom} where the risk of committing type I errors as more comparisons are made in statistical tests is analyzed, it is interesting that the conclusion they reach is that it is necessary to test with a number significantly higher than the one recommended by the ITU. In our case, this study supports the choice of 50 people, a number significantly higher than recommended.

%%Por otro lado, diversos estudios han sido consultados. En el paper de [3] donde se analiza el riesgo de cometer errores tipo I a medida que se realizan mas rcomparaciones en tests estadisticos. Es interesante que la conclusion a la que llegan es que es necesario hacer pruebas con un numero sensiblemente superior al que recomienda la ITU. En nuestro caso, este estudio avala la eleccion de 50 personas, numero sensiblemente superior al recomendado.
%%
%%
%El paper [4] entra en mucha profundidad sobre paliar posibles sesgos humanos que añaden ruido a las valoraciones subjetivas. Este interesante trabajo puede tomarse como linea futura sobre nuestros datos recogidos. Las conclusiones hechas en nuestro estudio han asumido que el elevado numero de participantes ha disminuido en gran medida el ruido en las valoraciones.\\
The paper~\cite{2017-Li} goes into great depth about alleviating possible human biases that add noise to subjective assessments. This interesting work can be taken as a future workline on our collected data. Conclusions made in our study have assumed that the higher number of participants has greatly reduced the noise in the assessments.

%Muchos mas documentos tratan sobre estos temas pero quiza algo tan simple como cuestionarse la propia estructura de lo que se esta evaluando pueda escaparse. Se ha asumido que una calidad objetivamente mejor tiene una correlacion inmediata con una mejora subjetiva. Esto generalmente es cierto, pero se puede matizar. Puede ser dependiente del dispositivo, del contenido y de la persona. Es en este punto donde este paper entra de forma transversal a todos los trabajos relativos a este tema.
Many more documents~\cite{2015-Janowski},~\cite{2016-Krasula},~\cite{2018-Narwaria} deal with these issues but perhaps something as simple as questioning the very structure of what is being evaluated might be ignored. It has been assumed that an objectively better quality has an immediate correlation with a subjective improvement. This is generally true, but it can be qualified. It can be device, content and person dependent. It is in this point where this paper enters transversally into all the work related to this topic.

\section{Test Features}
\label{sec:test_features}

In this section we summarize all the procedures and considerations that have been taken into account during the different test phases. Many of them are based on the operational guidelines included in Recommendations BT.500~\cite{2012-itu}, P.910~\cite{2008-itu} and P.913~\cite{2016-itu}.

\subsection{Methodology}
\label{subsec:methodology}

The conducted tests consisted in sequentially presenting all the PVSs in random order to all subjects in all the considered devices. The subjects were asked to rate on paper each sequence immediately after its visualization. To enable it, a four-second grey sequence was included between every two consecutive PVSs. The test method followed in the assessment is the Absolute Category Rating (ACR), where subjects have five possible answers to indicate the quality just perceived: "Excellent", "Good", "Fair", "Poor" and "Bad". 

Before the test starts, the test designer reads out loud the guidelines on how the subjects should perform the tests. Next, previous to the use of each device, a rather short training session is carried out. Subjects are presented with two possible conditions: the highest and lowest qualities that will be shown, i.e., 1080p and 270p. To that end, a sixth sequence is used. In this way, subjects are more aware of the scale of qualities that they will encounter and rate the sequences accordingly. Furthermore, this training session was also used to allow the subject to tune the audio level amplitude as he/she pleases.

\subsection{Material}
\label{subsec:material}

The test material consists of five different Source sequences (SRCs), all of them with the same resolution (4K Ultra High Definition (4K UHD), that is, 2160p), captured at a frame rate of 50 or 60 fps and of very high quality. All SRCs are 10 seconds long. They were chosen to include a representative, varied and habitual-to-observers set of contents: movies, TV series, documentaries and cartoons. This material can be characterized in terms of spatial and temporal complexity. In this work, they are computed respectively in terms of the spatial information (SI) and the temporal information (TI) of the sequence~\cite{2015-Janowski},~\cite{2008-itu}. The complexity is computed as the average of the standard deviation of the SI or TI for every frame. Table~\ref{tab:material_complexity} includes the spatial and temporal complexity of every SRC. As can be observed, the selected SRCs are of a wide range of complexities. Furthermore, all the SRCs include audio that matches the type of content presented.

\begin{table}[!t]
\caption{\label{tab:material_complexity}SRCs' spatial and temporal complexity}
\centering
\begin{tabular}{c|c|c}
Content & Spatial Complexity & Temporal Complexity\\
\hline
Game & 42.2 & 14.2\\
Venice & 80.7 & 11.7\\
Football & 62.7 & 18.0\\
India & 70.7 & 13.3\\
Skate & 38.5 & 16.4
\end{tabular}
\end{table}

We have considered the following Hypothetical Reference Circuits (HRCs): 
\begin{itemize}
\item HRC1: 270p. 
\item HRC2: 360p. 
\item HRC3: 540p. 
\item HRC4: 720p. 
\item HRC5: 1080p. 
\end{itemize}
all of them encoded with AVC at 30 fps, a target bitrate of 15~Mbps and progressive scan. That is, they only differ in the image resolution. Considering the devices targeted in the study, all HRC were selected to have a 16:9 aspect ratio. The particulary high bitrate value was selected to guarantee that the quality presented to the subjects only depended on the resolution of the video and not on any encoding parameter values.

So, finally, we have a total of 25 Processed Video Sequences (PVSs) under evaluation.

\subsection{Mobile Devices}
\label{subsec:device}

We have selected three smartphones with very different screen sizes and native resolutions, and one tablet. In order to prevent the tablet from distorting subjects' ratings, it was always shown either first or last, but never in between. In particular, the handheld devices chosen were the following:

\begin{itemize}
  \item LG-P720: with a 4.3-inch screen and a resolution of 480x800 pixels.
  \item Samsung Galaxy A3 (2017): with a 4.7-inch screen and a resolution of 720x1280 pixels.
  \item Samsung Galaxy S7: with a 5.1-inch screen and a resolution of 1440x2560 pixels.
  \item iPad Air 2: with a 9.7-inch screen and a resolution of 1536x2048 pixels.
\end{itemize}

\subsection{Environment}
\label{subsec:environment}

The environment of the room where the tests were conducted was set as comfortable as possible, with the aim of simulating to the greatest extent possible the habitual place in which the subjects usually use their mobile devices to enjoy audiovisual content. In addition, the light ranges were controlled so as to meet recommended values~\cite{2008-itu}. The specific numerical values of the brightness present in the room are presented in the table \ref{tab:brightness_room}.

\begin{table}[!t]
\caption{\label{tab:brightness_room} Brightness of the room where tests took place.}
\centering
\begin{tabular}{c|c|c|c|c}
\ & Front & Left & Right & Ceiling\\
\hline
Brightness (Lx) & 69.6 & 80.6 & 94.9 & 120.6
\end{tabular}
\end{table}

\section{Test Results}
\label{sec:results}

\begin{figure}[!t]
\centering
\includegraphics[width=\columnwidth]{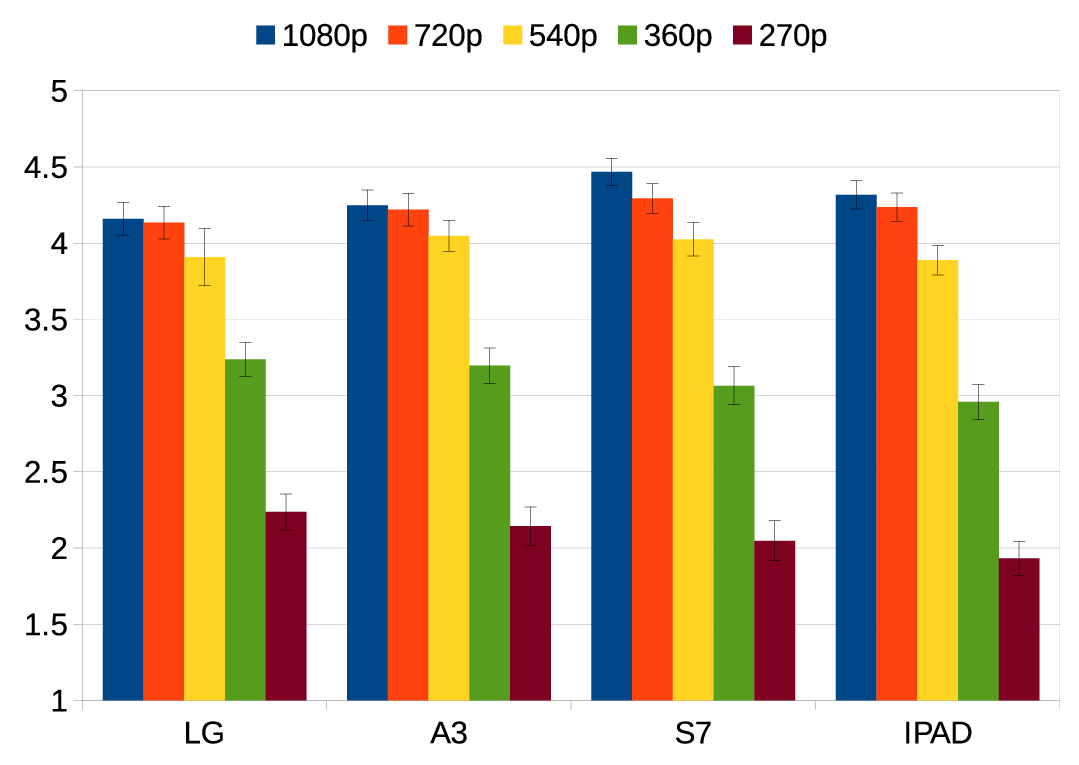}
\caption{Average score regarding quality and device.}
\label{averagequality}
\end{figure}

50 non-expert people participated in the test. Women and men on par, aged between 18 and 30 years, most of whom use mobile devices frequently to watch audiovisual content.

In Figure~\ref{averagequality}, depicts the mean opinion scores (MOSs) per resolution and mobile device. It results from the aggregation of the data obtained for the five SRCs. It includes 95\% confidence intervals (CIs). We can see that, as expected, the MOS increases with the number of pixels of the image in all the cases. However, some CIs clearly overlap. The main conclusion is then clear: subjects perceived no statistically significant difference between resolutions 1080p and 720p, regardless of the device. Furthermore, in some of the devices, they did not even perceive any statistically significant difference between resolutions 720p and 540p, but this result cannot be extended to all the condidered devices. Therefore, the equivalent resolution for handheld devices is 720p.

\begin{figure}[!t]
\centering
\includegraphics[width=\columnwidth]{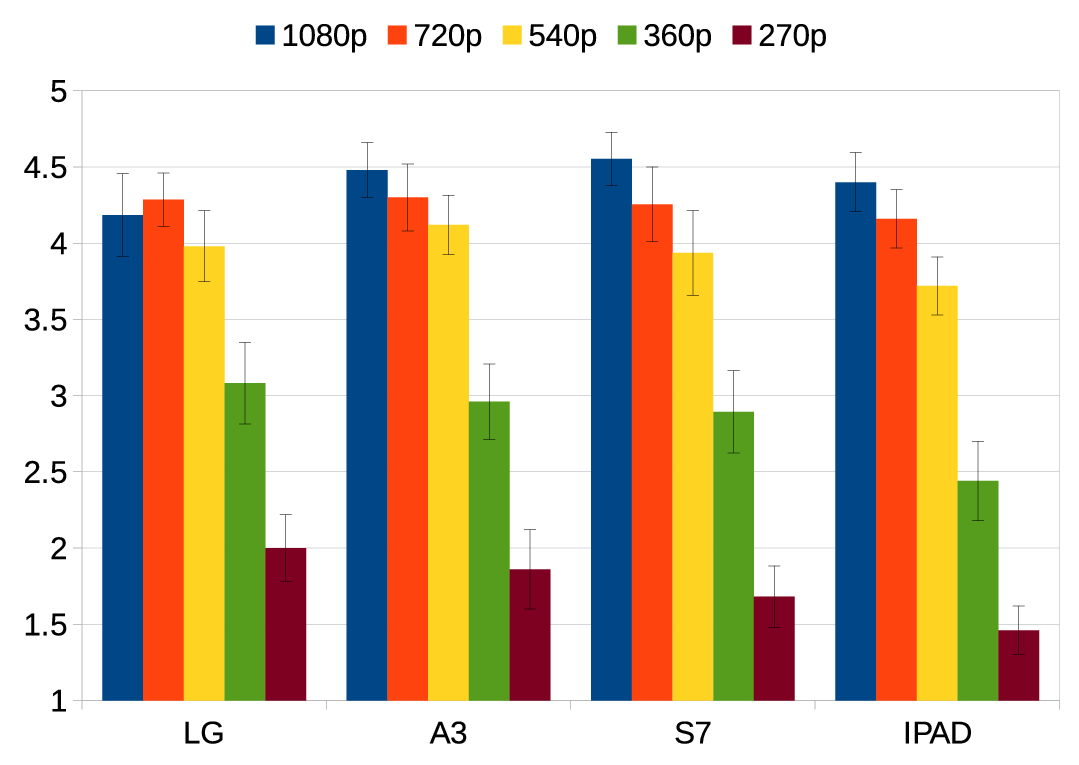}
\caption{Average score in "India" content regarding quality and device.}
\label{averageindia}
\end{figure}

\begin{figure}[!t]
\centering
\includegraphics[width=\columnwidth]{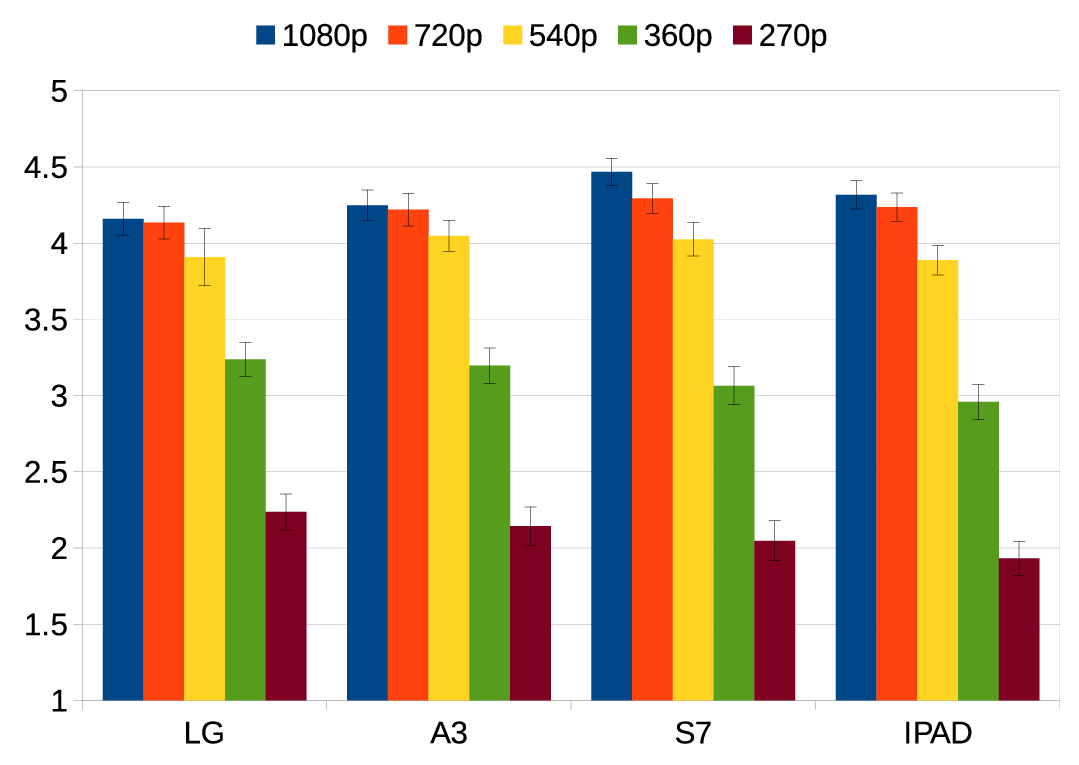}
\caption{Average score  in "Skate" content regarding quality and device.}
\label{averageskate}
\end{figure}

\begin{figure}[!t]
\centering
\includegraphics[width=\columnwidth]{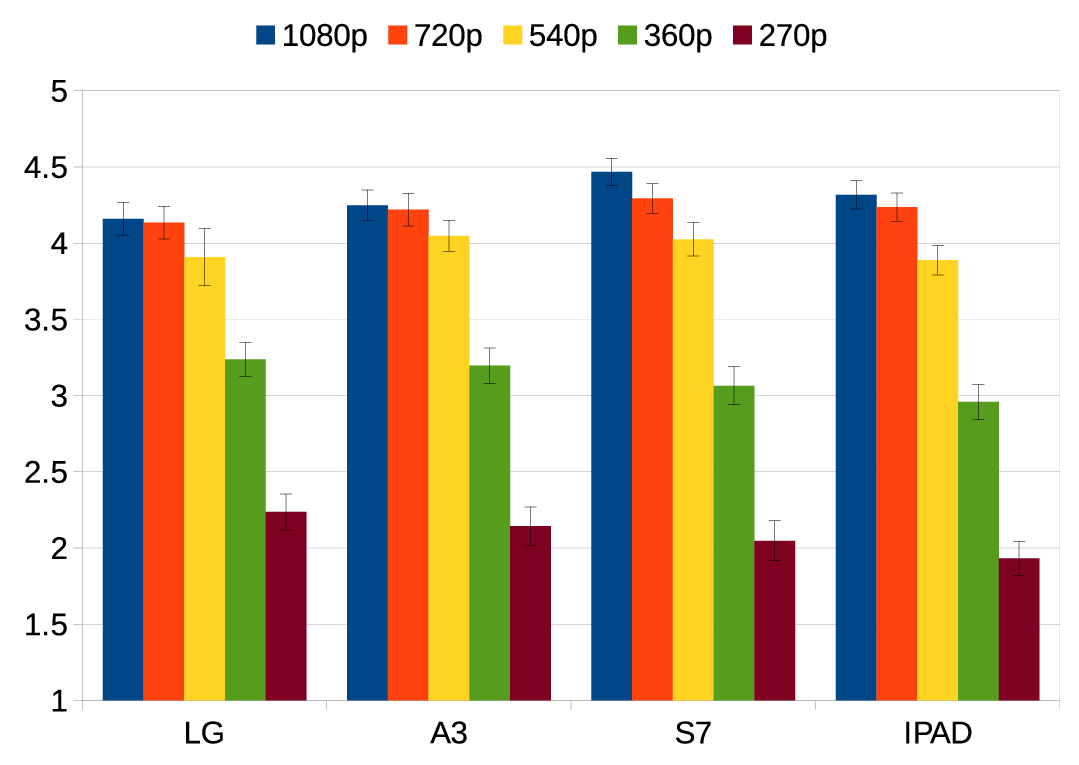}
\caption{Average score in "Football" content regarding quality and device.}
\label{averagefootball}
\end{figure}

\begin{figure}[!t]
\centering
\includegraphics[width=\columnwidth]{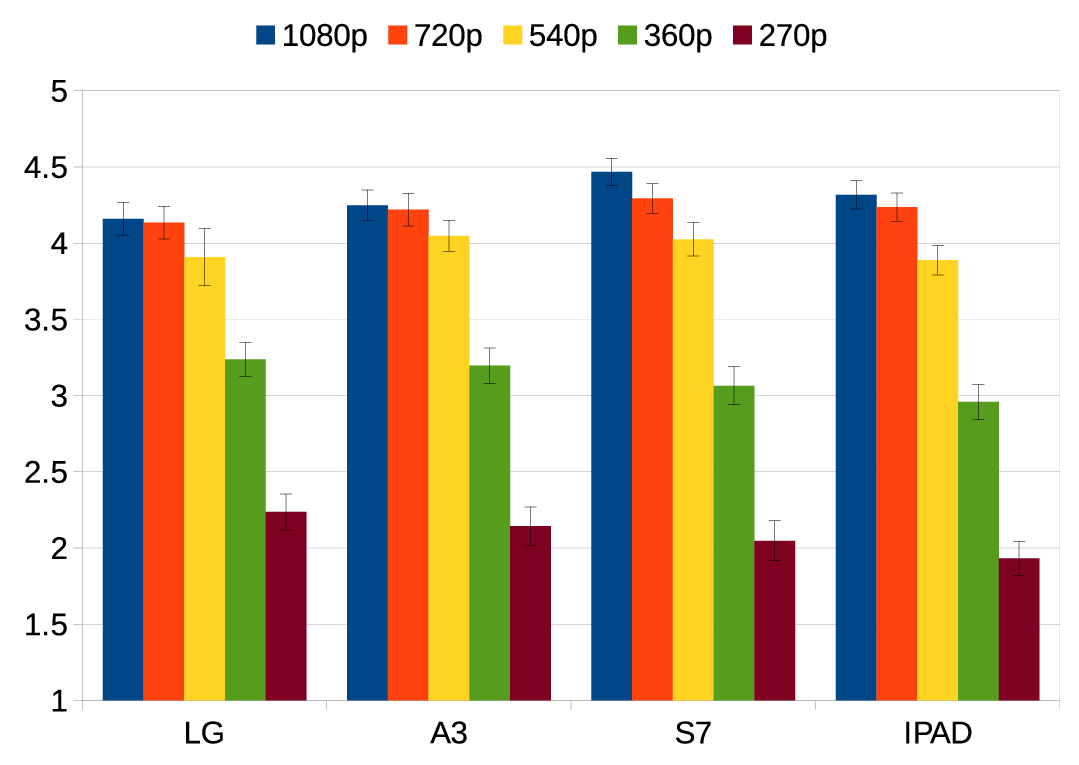}
\caption{Average score in "Game" content regarding quality and device.}
\label{averagegame}
\end{figure}

\begin{figure}[!t]
\centering
\includegraphics[width=\columnwidth]{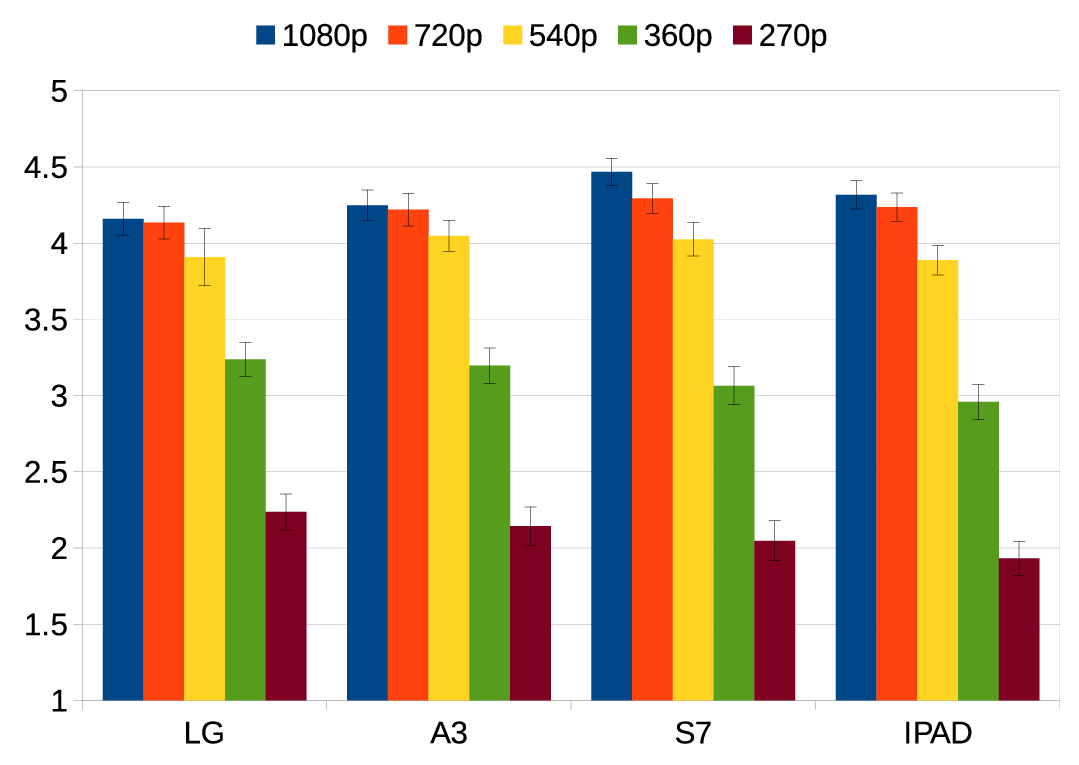}
\caption{Average score in "Venice" content regarding quality and device.}
\label{averagevenecia}
\end{figure}

Figures \ref{averageindia}, \ref{averageskate}, \ref{averagefootball}, \ref{averagegame}, \ref{averagevenecia} compare the average values that were granted to each of the resolutions according to each mobile device and SRC. The format is the same as for the previous graphic. In this case, the information adopts a more granular form. Some interesting conclusions are the fact that subjects tend to rate higher the lower resolutions on poor quality devices, while on good quality devices they tend to be rated lower. On the contrary, the best qualities in good devices are rated higher than in the worst devices. This shows that the dynamic range in low-end mobiles is lower, and therefore the demands are more lax. In high-end mobiles, the exigency is greater. Another conclusion is the fact that the lower resolutions in "India" SRC were rated slightly worse than the rest of SRCs. One possible answer has to do with the complexity, both spatial and temporal that the sequence has. At low resolutions, small objects may have disappeared completely and might had caused a superior displeasure to the subjects.

%The results seem logical. The worse the mobile device is, the more the value of all contents tends to be homogenized. As the characteristics of the terminal improve, higher differences can be found. Subjects statistically agree on considering that the content of greater spatial and temporal complexity to be better than the rest. Finally, the content "skate" seems to present big differences depending on the device. The conclusions are difficult to draw, because the tastes of the subject can play an important role.

\begin{figure}[!t]
\centering
\includegraphics[width=\columnwidth]{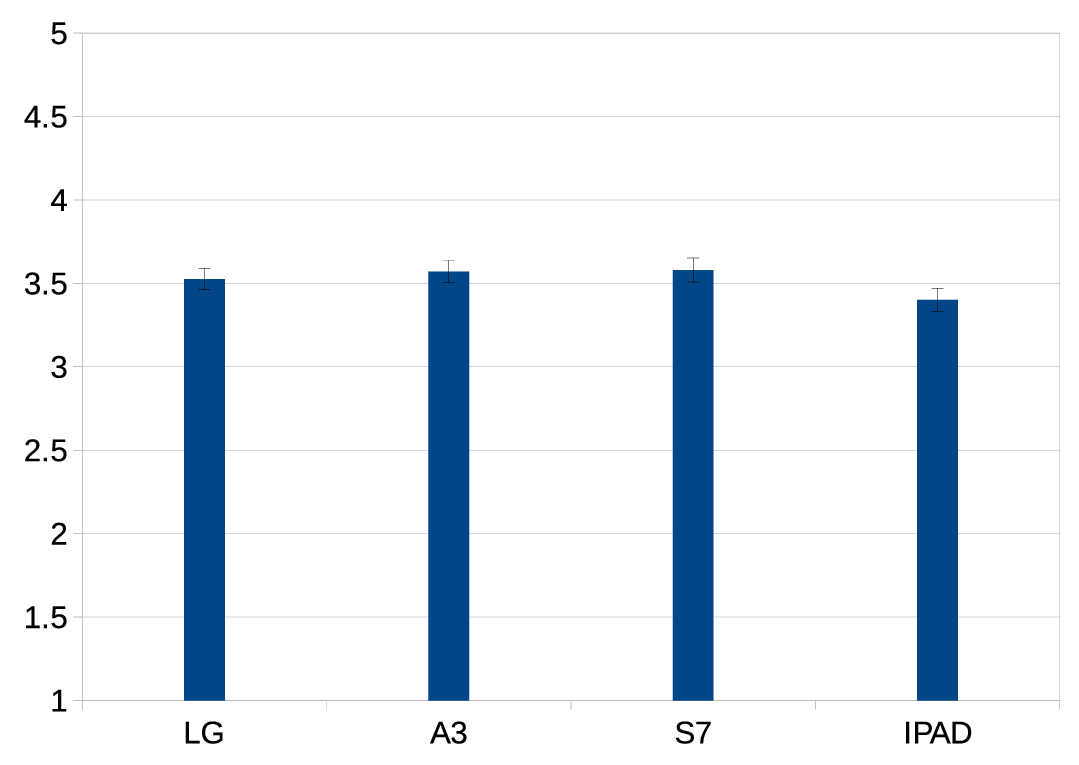}
\caption{Average score regarding quality, content and device.}
\label{averagefull}
\end{figure}

Figure \ref{averagefull} compares the average values that were given to each of the devices, without attending to the content or quality. The iPad is rated as the worst of the four devices. The conclusion is that the artifacts that the image can produce when oversampling are more visible in devices with larger screens.

%Figures from 19 to 24 compare the average values given to each content and its quality according to the mobile device. They are the most significant graphics to our goals. The following important conclusions can be drawn. The 270p quality is always considered to be very bad (taking into account that although the graphs range from 0 to 5, the minimum score that could be given in the test is 1). In addition, in some cases, very similar values were obtained for the three best qualities (very close to 5, in fact).

Additionally, a parallel study has been made in which the average score of a particular resolution has been compared with what was visualized immediately before. The conclusion is that the results are totally noisy. That is, there is no linear relationship between the results. This is in fact opposite good result, as it means that there has been no bias in this sense and that the randomization made for the presentation of the samples is valid.

\section{Conclusion}
\label{sec:conclusion}

The fact of delimiting the subjects between the best quality and the worse quality that are going to see has improved  the  comprehension  of  which  it  was  tried  to  be  evaluated  in  the  test.  The  questions  on  this character  have  been  considerably  reduced.  On  the  other  hand,  eliminating  this  degree  of  freedom  also eliminates  the  possibility  of  making  quality  comparisons  between  devices,  since  the  testers  know  what they can expect from a low-end mobile phone, such as the LG, as well as what they can expect from an iPad.  However,  according  to  the  graph  that  compares only  devices,  it  is  shown  that  even  so,  the  subject continues to rate the iPad as worse.
It has been proven that small breaks, even less than a minute, helped the tester to make the assessment more enjoyable. In this test the breaks were made every time 
they changed mobile devices.
It makes little sense to send a 1080p resolution to mobile devices, even if the resolution of the screen is greater than that one. It is  demonstrated that only in response to this  criterion, there is no difference in perception with respect to 720p. On the  other hand, sending images at 540p compared to 720p can be studied as an interesting possibility, since the perception is reduced little and a great bandwidth is saved. 
The options clearly to discard are 360p and 270p.

\section*{Acknowledgment}
This work has been partially supported by the Ministerio de Ciencia, Innovaci\'on y Universidades (AEI/FEDER) of the Spanish Government under projects IDI-20170572 (BUSTOP) and TEC2016-75981 (IVME).

\ifCLASSOPTIONcaptionsoff
  \newpage
\fi

\bibliographystyle{IEEEtran}
\bibliography{refs_spl2019}

\end{document}